\documentclass[twocolumn,aps,prd,showpacs,%
nofootinbib,eqsecnum]%
{revtex4}

\usepackage{amsmath,amssymb}
\usepackage{graphicx}

\newlength{\alexwidth}

\newcommand{\avtop}[2]{\genfrac{}{}{0pt}{}{#1}{#2}}

\renewcommand{\Re}{{\rm Re\thinspace}}
\renewcommand{\Im}{{\rm Im\thinspace}}

\allowdisplaybreaks
\begin{document}

\title{CP violation in unpolarized
$e^+ e^-\to$ charginos at one loop level}

\author{P.~Osland$^a$}
\email[Email address: ]{Per.Osland@ift.uib.no}
\author{A.~Vereshagin$^{a,b}$}
\email[Email address: ]{Alexander.Vereshagin@ift.uib.no}
\affiliation{
$^a$Department of Physics and Technology,
    Postboks 7803,
    N-5020 Bergen,
    Norway \\
$^b$Theor.\ Phys.\ Dept., Institute of Physics,
    St.Petersburg State University,
    St.Petersburg, Petrodvoretz,
    198504, Russia
}

\date{June 2007}

\begin{abstract}
We study CP violation in 
$e^+ e^- \to \tilde\chi_i^+ \tilde\chi_j^-$ in the framework of the
MSSM. Though the cross section of this process is CP-even at the tree
level even for polarized electron-positron beams, we show that it
contains a CP-odd part at the one loop order and there are CP-odd
observables that can in principle be measured even using
{\em unpolarized} electron-positron beams. The relevant diagram
calculations are briefly discussed and the results of selected
(box) diagram computations are shown.
\end{abstract}

\pacs{11.30.Er, 12.60.Jv, 14.80.Ly}

\maketitle

\section{Introduction}

The complex phases of the higgsino and gaugino mass parameters in the
Minimal Supersymmetric Standard Model (MSSM 
\cite{MSSM,DreesBook}) allow for CP violation at low orders of
perturbation theory, without invoking the Cabibbo--Kobayashi--Maskawa
matrix or the Higgs sector. If the phases are significant, one may
expect experimental evidence of CP violation that does not fit the
explanation within the (non-supersymmetric) Standard Model, leave
alone the consequences for CP-conserving processes. It has long been
known that these phases, if 
${\cal O}(1)$, could lead to values for the electron and neutron
electric dipole moments that would violate the experimental bounds
unless the superparticles had masses of 
${\cal O}(\text{TeV})$ or higher
\cite{Ellis:1982tk}. However, it has recently been realized that
there could well be cancellations among various contributions to such
CP-violating effects
\cite{Kizukuri:1992nj,MSSM_phases}, such that the experimental
constraints are respected, even with some phases of 
${\cal O}(1)$ and some superparticles light.

The couplings with potentially CP-violating phases affect many cross
sections and rates. However, the most informative way to study such
couplings would be in some CP-odd observable that would be accessible
in future experiments. In the light of the International Linear
Collider project 
\cite{Accomando:1997wt}, it is natural to consider the products of
$e^+ e^-$ annihilation. The chargino pair creation
\begin{equation}
e^+ + e^- \to \tilde\chi_i^+ + \tilde\chi_j^-
\label{eq:original-process-intro}
\end{equation}
then immediately comes to mind. At tree level the neutralino couplings
do not enter in the amplitude and the only CP-violating phase that
enters is $\phi_\mu$ due to the higgsino mass parameter 
\begin{equation} \label{Eq:mu-def}
\mu \equiv |\mu| e^{i \phi_\mu}. 
\end{equation}
This phase is indeed accessible at tree level if one in mixed events ($i\ne
j$) measures the transverse polarization of one of the charginos
\cite{Kizukuri:1993vh,Choi:1998ei,Bartl:2004vi}. There is also an additional
CP violating effect in chargino decays, at the one-loop level
\cite{Yang:2002am}.  However, if one does not consider the decay of a
final-state chargino, the tree-level cross section of the above process
conserves CP (in the $m_e = 0$ limit) \cite{Kizukuri:1993vh}, even if one
considers polarized electron-positron beams \cite{Bartl:2004xy}.

At the same time, there is no physical {\em symmetry}
which whould prohibit the cross section from acquiring a CP-odd
part: the result of \cite{Bartl:2004xy} is mainly dictated by the V-A
structure of the tree-level couplings (see the general discussion of the
effective form factors given in
\cite{Dass:1975mj,Ananthanarayan:2006yj}). Since by the very
construction MSSM is renormalizable and the tree-level cross section is CP
even, any non-vanishing CP-odd contribution should at one loop be {\em finite}
--- that is the logic of renormalization and that is why many regularization
problems \cite{Hollik:2002mv} drop out for this effect (see
Sec.~\ref{sec:loop-diagrams}).

Typically, to build a (scalar) CP-odd observable in a
\mbox{$2 \to 2$} process one has to employ spin (polarization) of one
of the particles in addition to the particle momenta, since any scalar
product of momenta is even under C and P
\cite{Ananthanarayan:2003wi,Branco}. However, the careful analysis in 
Sec.~\ref{sec:CP-transform} shows, that when the final chargino mass
indices are different,\footnote{We use a ``mass index'', taking
values 1 and 2, to distinguish the two chargino mass states.} 
their interchange should also be accounted
for and a CP-odd observable is easily constructed out of unpolarized
cross-sections. So, the CP-violation may in principle be observed in
the reaction
(\ref{eq:original-process-intro}) without any spin detection and with
unspecified polarization for the initial beams. This is the main
result of the present paper.

While the identified effect is radiatively induced, and thus of ${\cal
O}(\alpha)$, there could be enhancements due to factors $\tan\beta$ or
$\cot\beta$.  In any case, we think an independent CP-violating effect is
worth attention, if some kind of supersymmetry should be realized in nature.
In particular, it may provide information on whether the chargino sector
contains more than two mass states, and information on the neutralino sector,
including the phase of the U(1) gaugino mass parameter, $M_1$, via the
$W^\pm\tilde\chi_i^\mp\tilde\chi_k^0$ couplings.

Following many authors we work within the simplest version of
unconstrained MSSM making no assumptions about the symmetry breaking
mechanisms
\cite{Heinemeyer:2004gx}, neither do we impose any constraints on the
CP-violating phases. The R-parity and the lepton flavour violation is
not permitted, though, as noted in 
\cite{Bartl:2004xy}, the modification for less constrained models
can easily be done. Besides, just to simplify sample calculations we
assume that all slepton masses are large,%
\footnote{
One could refer to the parameter space area around the so-called
SPS 2 benchmark point
\cite{Allanach:2002nj}, but remember that the latter classification
assumes an mSUGRA breaking mechanism with no CP-violating phases.
}
and, of course, neglect everything proportional to the electron mass.
We do not calculate the one-loop cross section here, neither do we
give a review of the magnitude of the CP-odd observables in various
parameter points. Instead, we pick a specific parameter set that
allows us to neglect certain diagrams and show that the effect is
indeed non-zero at the one-loop order. 

\section{CP transformation of the cross section and the CP-odd
         observable}
\label{sec:CP-transform}

Let us consider chargino production in 
$e^+ e^-$ annihilation allowing for polarized initial beams:
\begin{equation}
e^+(p_1, P_+) + e^-(p_2, P_-)
\to \tilde\chi_i^+(k_1) + \tilde\chi_j^-(k_2),
\label{eq:original-process}
\end{equation}
where 
$P^\mu_\pm$ are the positron and electron polarization four-vectors
\cite{Bjorken&Drell} (see also \cite{Haber:1994pe}). 
The crucial point here is that for
$i \neq j$ the charginos do not form a particle-antiparticle pair.
Hence, while the initial state (for suitably chosen polarizations,
$P_+ \leftrightarrow P_-$) is in the c.m.\ frame odd under charge
conjugation, the final state has no such symmetry. We shall take a
closer look at this. 

The C and P unitary operators act in Fock space and transform the
creation operators as 
\cite{Weinberg_book,FeinWein}:
${\rm P} a^\dag ({\boldsymbol p}, \sigma, n) {\rm P}^{-1} 
= \eta_n  a^\dag ( -{\boldsymbol p}, \sigma, n) ;$
${\rm C} a^\dag ({\boldsymbol p}, \sigma, n) {\rm C}^{-1} 
= \xi_n  a^\dag ({\boldsymbol p}, \sigma, n^c)$,
where 
$\eta$ and
$\xi$ are the intrinsic space inversion and charge-conjugation
phases (parities),
${\boldsymbol p}$ is the three-momentum, 
$\sigma$ labels spin components, while
$n$ and
$n^c$ refer to other quantum numbers%
\footnote{
Like charge, chargino mass index, etc. Following the most common
convention, we treat the  
{\em positive} chargino as particle, its antiparticle is, of course,
the negative chargino with the {\em same} mass (mass index).
}
for particle and antiparticle, respectively. Hence, under P, C,
and CP conjugation the $S$-matrix element
$\langle 
\tilde\chi^+_i ({\boldsymbol k}_1), \tilde\chi^-_j ({\boldsymbol k}_2)
| S | e^+ ({\boldsymbol p}_1, P_+), e^- ({\boldsymbol p_2}, P_-)
\rangle$
of the process
(\ref{eq:original-process}) gets transformed into (up to a phase
that does not affect the cross section):
\begin{alignat}{2}
&\text{P:} &\quad
&\langle 
\tilde\chi^+_i (-{\boldsymbol k}_1),
\tilde\chi^-_j (-{\boldsymbol k}_2)
| S | 
e^+ (-{\boldsymbol p}_1, P_+), 
e^- (-{\boldsymbol p}_2, P_-)
\rangle; \nonumber \\
&\text{C:} &\quad
&\langle
\tilde\chi^-_i ({\boldsymbol k}_1),
\tilde\chi^+_j ({\boldsymbol k}_2)
| S |
e^- ({\boldsymbol p}_1, P_+), 
e^+ ({\boldsymbol p}_2, P_-) 
\rangle; \nonumber \\
&\text{CP:} &\quad
&\langle 
\tilde\chi^+_j ( -{\boldsymbol k}_2), 
\tilde\chi^-_i ( -{\boldsymbol k}_1)
| S |
e^+ (- {\boldsymbol p}_2, P_-), 
e^- ( - {\boldsymbol p}_1, P_+)
\rangle.
\end{alignat}
Thus, the
cross section for the P-conjugated process can be obtained by the
change of sign of the particle three-momenta: 
${\boldsymbol p}_{1,2} \leftrightarrow -{\boldsymbol p}_{1,2}, \;
{\boldsymbol k}_{1,2} \leftrightarrow -{\boldsymbol k}_{1,2}$;
the C-conjugation amounts to the following substitution in the cross
section:
${\boldsymbol p}_1 \leftrightarrow {\boldsymbol p}_2, \;
{\boldsymbol k}_1 \leftrightarrow {\boldsymbol k}_2, \;
m_i \leftrightarrow m_j, \;
P_+ \leftrightarrow P_-$;
and the CP-transformation results in the change:
${\boldsymbol p}_1 \leftrightarrow -{\boldsymbol p}_2, \;
{\boldsymbol k}_1 \leftrightarrow -{\boldsymbol k}_2, \;
m_i \leftrightarrow m_j, \;
P_+ \leftrightarrow P_-$.

To find candidates for CP-sensitive observables, let us write the
cross section as
\[
d\sigma = d\sigma_0 
+ ({\rm terms\; linear\; in\; } |{\boldsymbol P}_\pm| 
)
+ (\ldots)|{\boldsymbol P}_-||{\boldsymbol P}_+|,
\]
where 
$d\sigma_0$ does not depend on polarization vectors and will be 
referred to as the
{\em unpolarized} part.
Due to Poincar\'{e} invariance 
$d\sigma_0$ may depend only on masses
$m_i, m_j$ and on two independent scalar variables, say, on
Mandelstam's
\mbox{$s \equiv (p_1 + p_2)^2$} and
\mbox{$t \equiv (p_1 - k_1)^2$}. The latter do not change under
C or P,
so the
CP-transformation for the unpolarized cross-section is reduced to the
interchange of the masses in the resulting formula%
\footnote{
Of course, the coupling constants at vertices with charginos should be
considered as functions of the chargino masses, or, better, the
mass indices
$i,j$.
}.
Therefore, for equal-mass fermions in the final state
($i = j$) the unpolarized cross section is always P-, 
C- and CP-even%
\footnote{
The famous forward-backward asymmetry term in the
{\em unpolarized} cross-section of, say,
$e^+e^- \to \mu^+\mu^-$ scattering, which is often referred to as
parity violating, in fact only indicates the presence of a parity
violating term in the interaction, the unpolarized cross-section 
itself being, of course, P-even.
}.
{\em In contrast, if the chargino species are different, CP-violating
terms can arise even in the unpolarized cross-section.} That is the 
effect we will consider here, so unless otherwise stated the
final-state chargino masses are taken non-equal. 

Calculations show that the tree-level cross section (polarized and
unpolarized) of the process (\ref{eq:original-process}) is CP even 
\cite{Bartl:2004xy}, but, as we shall see, CP-odd terms do arise in
the one-loop contributions. Therefore, a natural 
experimental observable to consider is the ratio
\begin{equation}
\frac{d\sigma_0^{\rm odd}}{d\sigma_0}\ ,
\label{eq:obvservable-definition}
\end{equation} 
where  
$d\sigma^{\rm odd}$ is the CP-odd part of the corresponding 
cross-section:
\begin{equation}
d\sigma_0^{\rm odd} = 
\frac{1}{2} \Bigl[ d\sigma_0 - d\sigma_0^{\rm CP} \Bigr] , \quad
d\sigma_0^{\rm CP} 
\equiv d\sigma_0 \Bigr|_{m_i \leftrightarrow m_j} .
\label{eq:sigma_odd_def}
\end{equation}

As just mentioned, the CP violation first enters at one loop, thus, to
estimate the effect one should caculate $d\sigma_0^{\rm odd}$ at the one-loop
level. On the other hand, in most of the kinematical regions far from any
resonance, one can expect (see, e.g,
\cite{Blank:2000uc,Diaz:2002rr,Fritzsche:2004nf,Oller:2005xg}) that the
tree-level gives a reasonable approximation to $d\sigma_0$ in the denominator
of Eq.~(\ref{eq:obvservable-definition}). So, we will deal only with the ratio
\begin{equation}
\frac{\left. d\sigma_0^{\rm odd} \right|_{\rm 1\; loop}
}{
\left. d\sigma_0 \right|_{\rm tree}}\ .
\label{eq:obvservable-loop/tree}
\end{equation}
In the following Sections we discuss the diagrams that (may)
contribute to this observable and provide some sample calculations.

\section{Diagrams}
\label{sec:Diagrams}	 

The MSSM spectrum and Lagrangian are reviewed by many authors (e.g.\
\cite{MSSM,DreesBook}), we use the Feynman rule collections of
\cite{Kileng,Rosiek:1989rs}. Following the latter article, we work in 
\mbox{'t Hooft}--Feynman gauge
\cite{'tHooft:1971fh,Fujikawa:1972fe}, though for more
involved loop calculations other gauge choices may be preferable
\cite{Hollik:2002mv,Non-linearGauge}. When drawing diagrams, we found
it convenient to indicate sparticles by double lines. Due to 
$R$-parity conservation,%
\footnote{
Recently the
R-parity non-conserving extensions of the MSSM started to attract
attention (see e.g.
\cite{Allanach:2003eb}), however here we do not consider these cases.
}
the total number of such lines attached to each vertex should be even.
Following
\cite{Denner:1992vz}, we do not indicate the (double) fermion line
direction for the neutralino and choose a convenient fermion flow for each
diagram.

\subsection{Tree diagrams}
\label{sec:tree-diagrams}

\begin{figure}[htp]
\begin{center}
\includegraphics[width=4.2cm]{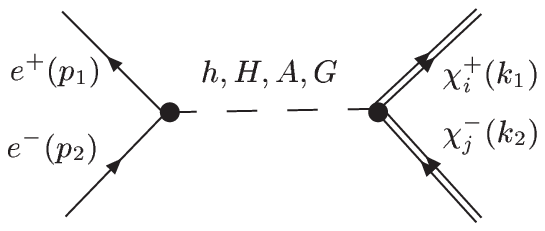}
\includegraphics[width=4.2cm]{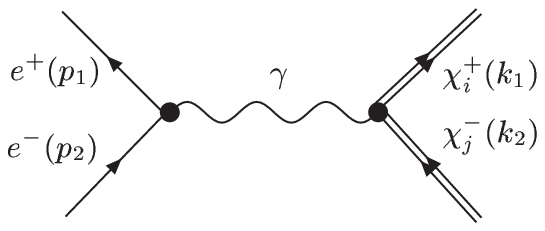}
\\
\hspace{0.0cm}
{\bf a.}\hspace*{3.5cm}
{\bf b.}\hspace*{3.5cm}
\\
\includegraphics[width=4.2cm]{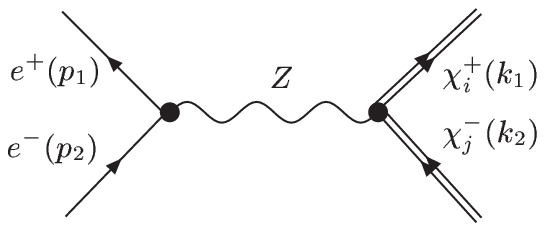} 
\includegraphics[height=3.5cm]{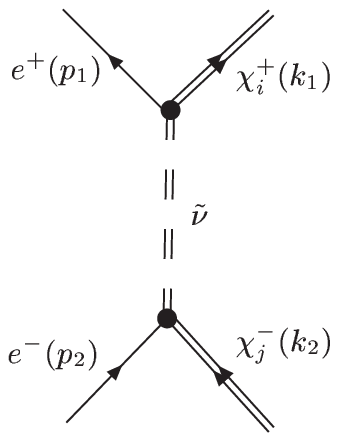}
\\
\hspace{0.0cm}
{\bf c.}\hspace*{4.0cm}
{\bf d.}\hspace*{2.5cm}
\caption{
Tree-level diagrams: superpartners of ordinary particles are pictured
by double lines.
\label{fig:tree} }
\end{center}
\end{figure}

We need the tree-level cross section to normalize the observable
(\ref{eq:obvservable-loop/tree}). The graphs contributing to the tree
amplitude
${\cal M}_{\rm tree}$ are drawn%
\footnote{
Diagrams are drawn by 
{\em JaxoDraw} tools \cite{Vermaseren:1994je}.
}
in
Fig.~\ref{fig:tree}. They are:
$s$-channel Higgs (and the unphysical Goldstone), photon and
$Z$ exchanges, and
$t$-channel sneutrino exchange.

The Higgs (Goldstone) exchanges can be dropped since their couplings
are proportional to 
$m_e$, while
$\gamma$ exchange is absent since the final-state charginos have
different masses and there is no non-diagonal coupling with the photon
in the MSSM (this is a requirement of gauge invariance and
renormalizability). Finally, to make sample loop calculations
simpler, we assume that all sleptons are heavy and, hence, only the 
$Z$-exchange contributes at tree level.

The differential cross section (in the c.m. system) is
\begin{equation}
\frac{d\sigma}{d\Omega}
= \frac{ \beta }{64 \pi^2 s} |{\cal M}|^2\ , 
\quad \beta \equiv 
\frac{|{\boldsymbol p}_{\rm out}|}{|{\boldsymbol p}_{\rm in}|}\ ,
\label{eq:dsigma_domega}
\end{equation}
and the direct calculation for unpolarized
$Z$-exchange amplitude gives (cf. 
\cite{Bartl:2004xy}):
\begin{eqnarray}
\lefteqn{|{\cal M}_{Z,\text{ tree}}|^2} 
& &
\nonumber \\
& = & 
\chi^2 \Bigl( (g_V^2 + g_A^2)
\bigl\{
|G_V|^2 [{\cal A} -2(m_i - m_j)^2/s]
\nonumber \\
&&
+ |G_A|^2 [{\cal A} - 2(m_i+m_j)^2/s]
\bigr\}
\nonumber \\
&&
- 4 g_V g_A (G_V^\ast G_A + G_V G_A^\ast) \beta\cos\theta
\Bigr),
\label{eq:tree-crosse-section-unpolarized}
\end{eqnarray}
where 
$s = (p_1 + p_2)^2$, 
$m_i$, $m_j$ are the chargino massess,
$\theta$ is the scattering angle, and
\[
\chi = \left(\frac{g}{4 \cos\theta_W}\right)^2\, \frac{s}{s-M_Z^2} ,
\quad
{\cal A} =2-\beta^2\sin^2\theta . 
\]
The 
$Zee$ (reduced) couplings are:
$g_V = 1 - 4\sin^2\theta_W$, $g_A = -1$,
and we use
$G_V \equiv G_{V\, j,i}$ and $G_A \equiv G_{A\, j,i}$ to abbreviate
the $Z\chi\chi$ coupling constants:
\begin{eqnarray}
L_{Z\chi\chi} 
& = & \frac{g}{4 \cos\theta_W} \bar{\Psi}_{\chi_j} \gamma^\rho
\Bigl\{
\Bigl[ 2\delta_{kj}\cos{2\theta_W} 
\nonumber
\\
& &
+ U_{k1}U_{1j}^\dagger +
V_{j1}V_{1k}^\dagger \Bigr]
\nonumber
\\
& &
+ \gamma^5 \Bigl[ U_{k1}U_{1j}^\dagger - V_{j1}V_{1k}^\dagger\Bigr]
\Bigr\}
\Psi_{\chi_k} Z_\rho
\nonumber
\\
& \equiv & \frac{g}{4 \cos\theta_W} \bar{\Psi}_{\chi_j} \gamma^\rho
\Bigl\{ G_{V\, k,j} 
\nonumber
\\
& &
+ \gamma^5 G_{A\, k,j} \Bigr\}
\Psi_{\chi_k} Z_\rho 
\label{L:GCN-2}
\end{eqnarray}
(note, that the first mass index of
$G_{V\, k,j}$ and 
$G_{A\, k,j}$ refers to the mass of the 
{\em annihilated particle}, which is the
{\em positive} chargino). The matrices 
$U$ and 
$V$ diagonalize the chargino mass matrix
$M_{\chi}$:
\begin{align}
M_{\chi} &=
\left(
\begin{array}{cc}
M_2 & \sqrt{2} m_W \sin\beta \\
\sqrt{2} m_W \cos\beta & \mu
\end{array}
\right)\nonumber \\
U^\ast M_{\chi} V^\dagger &= 
\left(
\begin{array}{cc}
m_{\chi_1} & 0\\
0 & m_{\chi_2} 
\end{array}
\right),\quad 0 < m_{\chi_1} < m_{\chi_2} .\
\label{eq:mass_matr_chargino}
\end{align}
The SU(2) gaugino mass parameter 
$M_2$ can always be chosen real, while 
$\mu$ (as well as the U(1) gaugino mass
parameter 
$M_1$ appearing in the neutralino mass matrix)
is in general complex quantity. 


According to 
Sec.~\ref{sec:CP-transform}, 
under the CP-transformation
$G_{V (A)\, j,i} \leftrightarrow G_{V (A)\, i,j} ,
\ m_i \leftrightarrow m_j$.
On the other hand, the hermiticity of the Lagrangian enforces the
relation
\begin{equation}
G_{V (A)\, j,i} = G^\ast_{V (A)\, i,j},
\label{eq:hermiticity_Z_chi_chi}
\end{equation}
so 
Eq.~(\ref{eq:tree-crosse-section-unpolarized}) is clearly CP-even.

If sneutrino exchange is not neglected, the cross section consist of
the squared graphs terms
(Fig.~\ref{fig:tree} {\bf c}, {\bf d}) and an interference term. Each
of them turns out to be CP-even
\cite{Bartl:2004xy}.

\subsection{Loops}
\label{sec:loop-diagrams}

The complete list of one-loop (prototype) graphs contributing to the
cross section can be found in
\cite{Diaz:2002rr}. The fact
that the tree-level cross section is CP even makes it evident that
many of the 
$d\sigma_0$ one-loop corrections cancel in
$d\sigma_0^{\rm odd}$
(\ref{eq:sigma_odd_def}), the numerator of
(\ref{eq:obvservable-loop/tree}). Indeed, the external wave function
renormalization is multiplicative, the propagator corrections result
just in a propagator mass shift,%
\footnote{
It is a bit more tricky if one sticks to precise one-loop order and does
not allow for the Dyson resummation in the propagators. Then each of
the tree graphs 
(Fig.~\ref{fig:tree}) acquires different functional (CP-even)
multiplier and the structure of the tree-level result
\cite{Bartl:2004xy} ensures that CP-odd terms cannot arise. We do
not demonstrate it here as we discard the sneutrino exchange graph,
and, therefore, will get a multiplicative correction anyway.
}
therefore we do not need to calculate the two-point functions and,
hence, neither Faddeev--Popov ghosts nor coloured particles will be
involved. In other words, there are only two types of one-loop
corrections that may contribute to
$d\sigma_0^{\rm odd}$: box diagrams and the tree diagrams from
Fig.~\ref{fig:tree} with a triangle loop instead of one of the vertices.
Before we take a closer look at the box diagrams (we do not compute the
triangle vertex corrections here), it is necessary to say a couple of
words about the ultraviolet and infrared behaviour of
$d\sigma_0^{\rm odd}$ at the one-loop order.

As mentioned in the Introduction,
$d\sigma_0^{\rm odd}$ must be UV finite, since it vanishes at tree
level: otherwise the counterterms required would mirror the tree level
CP-odd contribution. So, no infinite (UV-divergent) counter\-terms are
required. In fact, one can also see that any 
{\em finite} counterterm just results in corrections to the
tree level vertices in   
Fig.~\ref{fig:tree}. In particular, in
Eq.~(\ref{eq:tree-crosse-section-unpolarized}) only
$g_{V,A}$ and
$G_{V,A}$ may get modified, and, since 
Eq.~(\ref{eq:hermiticity_Z_chi_chi}) should always hold, the result
will still be CP-even and no contributions to $d\sigma_0^{\rm odd}$
will arise. One can easily check that the unpolarized cross section
with sneutrino exchange will also be unaffected by counterterms. This
relates also to finite counterterms which may be required to restore
the symmetries violated by regularization%
\footnote{
There are, however, no such simple arguments for
{\em polarized} amplitudes, as one of the potential CP-odd terms
in this case is cancelled due to the tree level SUSY relation between
chargino and sneutrino couplings
\cite{Bartl:2004xy}. The symmetry-restoring counter\-terms may, in
general, violate this relation and therefore can give an additional 
CP-odd term. We shall not discuss it here.
}
in the so-called algebraic renormalization approach
\cite{Hollik:2002mv}. So, at least for the unpolarized cross section,
we should not worry about the renormalization scheme (we assume
that the on-shell normalization conditions are used) and the
standard dimensional regularization will be adequate at the one loop order:
all divergent pieces must cancel in
$d\sigma_0^{\rm odd}$.

The situation with infrared (IR)
finiteness is slightly more complicated: there are many loops with
massless particles inside. However, according to
\cite{Hollik:2002mv}, all the IR singularities that appear at any loop
order in our amplitude are of the standard type, namely, they arise
due to the soft photons and cancel when real bremsstrahlung is
accounted for. On the other hand, the bremsstrahlung photon emission
from the tree diagram results in just an overall factor for the
corresponding amplitude.
Since the tree amplitude is CP-even, we conclude that 
$d\sigma_0^{\rm odd}$ is free of IR singularities.

Each possible box diagram turns out to be UV-finite just by power
counting. Since we assume heavy sleptons, any box with a slepton
line can be neglected. The only box diagrams that may contribute to
Eq.~(\ref{eq:obvservable-loop/tree}) in this limit are drawn in 
Fig.~\ref{fig:vgg}. Those are the only graphs whose contribution to
$d\sigma_0^{\rm odd}$ we shall evaluate numerically. Analytical
results for the coefficients of the box type Passarino--Veltman-like
functions presented in the next section ensure that the CP-odd
contribution from box diagrams can not be completely cancelled by graphs
with triangle loop corrections and, hence, the CP-violation is indeed
present in the unpolarized cross section.

\section{Numerical estimates}
\label{sec:Numbers}

Loop amplitudes are conveniently evaluated in terms of 
Passarino--Veltman functions 
\cite{Passarino:1978jh}. In
\cite{Diaz:2002rr} the cross section of the process
(\ref{eq:original-process-intro}) was parametrized
in terms of those functions and calculated in various parameter
points. However, the latter results were obtained assuming a
CP-invariant theory (real couplings) and (to make the results compact) 
the reduction to 
scalar Passarino--Veltman functions was not done. Since only the
scalar functions can be considered independent (differ from each other
by singularity pattern) we performed this reduction in our formulae.

\begin{figure}[htb]
\begin{center}
\includegraphics[width=4cm]{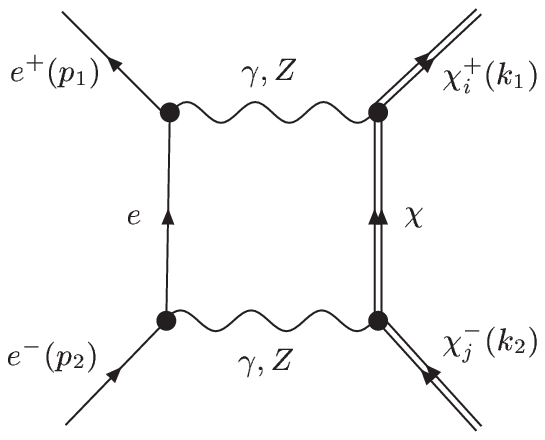} 
\includegraphics[width=4cm]{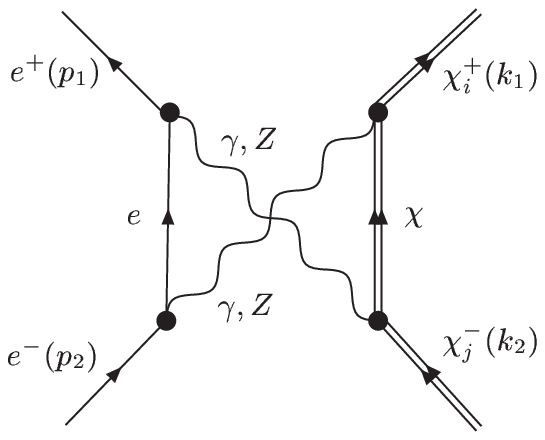} 
\includegraphics[width=4cm]{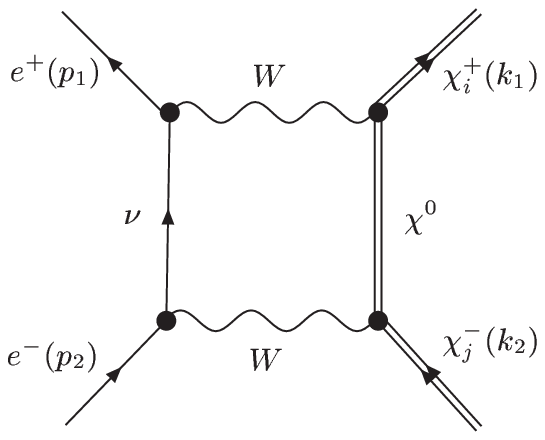} 
\caption{
Box diagrams without slepton lines: all chargino and neutralino
mass eigenstates contribute.
\label{fig:vgg}
}
\end{center}
\end{figure}

\begin{figure}[htb]
\begin{center}
\includegraphics[width=7cm]{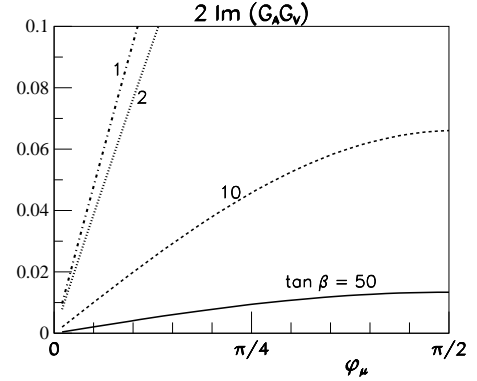}
\caption{
The couplings of
Eq.~(\ref{Eq:couplings}) vs.\ $\phi_\mu$ for various values of
$\tan\beta$. 
\label{fig:couplings} }
\end{center}
\end{figure}

At one loop order the cross section is defined by
Eq.~(\ref{eq:dsigma_domega}) with
\[
\left| {\cal M}^2 \right|_{\rm 1\, loop} 
= {\cal M}_{\rm tree}^\ast {\cal M}_{\rm 1\, loop} 
+ {\cal M}_{\rm tree} {\cal M}_{\rm 1\, loop}^\ast \ ,
\]
and, since we assume heavy sleptons, the tree amplitude
${\cal M}_{\rm tree}$ contains only the
$s$-channel 
$Z$-exchange graph of
Fig.~\ref{fig:tree}. Direct calculations
\cite{PhD} show that the CP-odd part of
$\left| {\cal M}^2 \right|_{\rm 1\, loop}$ accquire four-point
(``box'') integral contributions. In particular, for the $Z$-exchange
(uncrossed and crossed) box diagrams of
Fig.~\ref{fig:vgg}, after reduction to scalar integrals one obtains
(the subscript ``D'' refers to terms proportional to genuine
box diagram functions, as defined below):
\begin{eqnarray}
\lefteqn{
\left| {\cal M}^2 \right|_{\avtop{\rm CP-odd,}{\rm Z-box,\; D}}
}
& & 
\nonumber
\\
& = & 
\frac{1}{(2\pi)^4} 
2 \Re \Bigl[
\frac{i g^6 m_i m_j}{ (4 \cos\theta_W)^6 }  
(G_{A\, ij}G_{V\, ji} - G_{A\, ji}G_{V\, ij})
\nonumber
\\
& & 
\times
\bigl\{
 g_A(g_A^2 + 3 g_V^2) 
m_Z^2 (G_{V\, ii} I_{i; ji} + G_{V\, jj} I_{j; ji})
\nonumber
\\
& & \mbox{} \; 
+ g_V(3 g_A^2 + g_V^2) \bigl[
  (2 m_i^2 - m_Z^2 - 2t) G_{A\, ii} I_{i; ji}
\nonumber
\\
& & \mbox{} \; 
\quad \quad \quad
+ (2 m_j^2 - m_Z^2 - 2t) G_{A\, jj} I_{j; ji}
\bigr]
\nonumber
\\
& & \mbox{} \; 
+ g_A(g_A^2 + 3 g_V^2) 
m_Z^2 (G_{V\, ii} I^{\rm cr}_{i; ji} + G_{V\, jj} I^{\rm cr}_{j; ji})
\nonumber
\\
& & \mbox{} \; 
- g_V(3 g_A^2 + g_V^2) \bigl[
  (2 m_i^2 - m_Z^2 - 2u) G_{A\, ii} I^{\rm cr}_{i; ji}
\nonumber
\\
& & \mbox{} \; 
\quad \quad \quad
+ (2 m_j^2 - m_Z^2 - 2u) G_{A\, jj} I^{\rm cr}_{j; ji}
\bigr]
\bigr\}
\Bigr]. \label{Eq:res-Z-box}
\end{eqnarray}
Here, as above, $m_i$, $m_j$ are the chargino masses
($i,j = 1,2$),
and couplings are defined in
Sec.~\ref{sec:tree-diagrams}.
From the tree diagram there is a coupling $g_V$ or $g_A$ at the $Zee$
vertex, and a $G_{V\, ji}$ or $G_{A\, ji}$ at the $Z\chi_i\chi_j$
vertex, whereas the box diagrams contribute two $Zee$ couplings
($g_V^2$, $g_A^2$ or $g_Vg_A$), and two $Z\chi\chi$ vertices,
one of which will be diagonal in mass index ($G_{V\, ii}$, 
$G_{A\, ii}$, $G_{V\, jj}$ or $G_{A\, jj}$), and one will be
non-diagonal ($G_{V\, ij}$, $G_{A\, ij}$, $G_{V\, ji}$ or 
$G_{A\, ji}$). The two non-diagonal $Z\chi\chi$
couplings factor out as the combination
\begin{equation} \label{Eq:couplings}
G_{A\, ij}G_{V\, ji} - G_{A\, ji}G_{V\, ij}
=2i\,\Im G_{A\, ij}G_{V\, ji}.
\end{equation}
This quantity is shown in Fig.~\ref{fig:couplings}, for the 
set of parameters:
\begin{equation} \label{Eq:mass-params}
|\mu| = 300~\text{GeV}, \quad M_2 = 200~\text{GeV}.
\end{equation}
We note that the quantity (\ref{Eq:couplings}) increases with 
decreasing values of $\tan\beta$.

The integrals $I$ and $I^{\rm cr}$ of Eq.~(\ref{Eq:res-Z-box})
are the Passarino--Veltman scalar four-point functions
which correspond to ``normal'' and ``crossed'' box diagrams in
Fig.~\ref{fig:vgg}, respectively:
\begin{eqnarray*}
I_{k; ij} 
& \equiv & D (p_1, p_2, -k_2, -k_1, m_Z, 0, m_Z, m_{\chi_k}) 
\\
I_{k; ij}^\text{cr}
& \equiv & D (p_1, p_2, -k_1, -k_2, m_Z, 0, m_Z, m_{\chi_k})\ ,
\end{eqnarray*}
where, following
\cite{Passarino:1978jh},
\begin{eqnarray}
\lefteqn{
D (l_1, l_2, l_3, l_4, m_1, m_2, m_3, m_4)
}
& &
\nonumber
\\
& \equiv & 
\int {\rm d^4} q 
\bigl\{
(q^2 - m_1^2)[(q+l_1)^2 - m_2^2]
\nonumber
\\
& & 
\quad \quad \; \times [(q+l_1+l_2)^2 - m_3^2]
\nonumber
\\
& & 
\quad \quad \; \times [(q+l_1+l_2+l_3)^2 - m_4^2]
\bigr\}^{-1}
\end{eqnarray}
(in numerical calulations one may favour a more symmetric loop 
momentum assignment which permits consistency cross tests
\cite{PhD}). Analogous, though more cumbersome, pieces follow
from the box diagram with
$W$-exchange (the $D$-pieces of the
$\gamma Z$-exchange box diagrams cancel) and one can check that all 
these four-point integral contributions do not cancel each other. 
Besides, two- and three-point integrals (denoted $B$ and $C$ in
\cite{Passarino:1978jh}) also appear after reduction of tensor box
integrals stemming from the diagrams in
Fig.~\ref{fig:vgg}. What is essential, is that while the graphs with
triangle vertex corrections may contribute
$B$ and $C$ (and, possibly,
$A$ --- the one-point) functions to 
$d\sigma_0^{\rm odd}$, the 
$D$ function can never appear in triangle diagrams. As the function 
$D$ cannot be constructed out of 
$A$, $B$, $C$ integrals and rational functions, we may conclude that 
$d\sigma_0^{\rm odd}$ is non-zero at the one-loop order.

Even assuming heavy sleptons the total box diagram contribution to
$d\sigma_0^{\rm odd}$ is too awkward%
\footnote{
For many algebraic manipulations REDUCE and MATLAB packages were
used.
}
to provide here the complete formulae. Instead, to give an idea about
the orders of magnitude, we shall provide some plots. We stress once
more that the triangle loop corrections to the tree-level vertices are
not accounted for, therefore the numbers given are purely
illustrative. Below, the ratio
(\ref{eq:obvservable-loop/tree}) (with the amplitude
${\cal M}_{\rm 1\; loop}$ built solely of the diagrams in  
Fig.~\ref{fig:vgg}) is plotted as a function of the CP-violating phase
$\phi_\mu$ of Eq.~(\ref{Eq:mu-def}), 
and for simplicity the U(1) gaugino mass parameter appearing in the
neutralino mass matrix is taken to be real:
$M_1 = 250~\text{GeV}$ . 
The absolute values of the remaining chargino and
neutralino mass matrix parameters are given by 
Eq.~(\ref{Eq:mass-params}).

\begin{figure}[htb]
\begin{center}
\includegraphics[width=7cm]{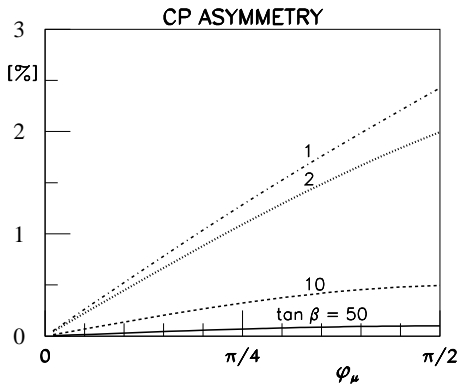}
\caption{
The ratio
(\ref{eq:obvservable-loop/tree}) at various values of
$\phi_\mu$ and
$\tan\beta$. The polar scattering angle is
$\theta = \pi/3$ while
$\sqrt{s} = 600\text{ GeV}$.
\label{fig:asymm_600} }
\end{center}
\end{figure}

We start this little numerical digression by showing in
Fig.~\ref{fig:asymm_600} the asymmetry resulting from the box
diagrams (for the numerical work, we used the {\tt LoopTools}
\cite{Hahn:1998yk,vanOldenborgh:1989wn} package), as a function of 
$\phi_\mu$, for $\sqrt{s}=600$~GeV, $\cos\theta=0.5$ and a few values 
of $\tan\beta$.
As anticipated, for $|\phi_\mu|\ll1$, the effect is linear in 
$\phi_\mu$. Also, we note that the shape of these curves
(i.e., dependence on $\phi_\mu$ and $\tan\beta$) is essentially
given by the coupling constants (\ref{Eq:couplings}) and shown in 
Fig.~\ref{fig:couplings}.

\begin{figure}[htb]
\begin{center}
\includegraphics[width=7cm]{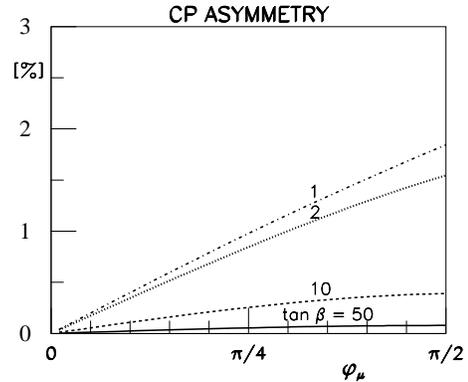}
\caption{
Same parameters as in 
Fig.~\ref{fig:asymm_600}, except
$\sqrt{s} = 800\text{ GeV}$.
\label{fig:asymm_800} }
\end{center}
\end{figure}

When the energy increases, the effect is reduced, as illustrated
in Fig.~\ref{fig:asymm_800}, where we show similar plots for
$\sqrt{s}=800$~GeV. The CP violation is related to the fact that 
the two charginos will have different velocities (due to
different masses). At high energies, the difference in masses plays
a lesser role.

\begin{figure}[htb]
\begin{center}
\includegraphics[width=7cm]{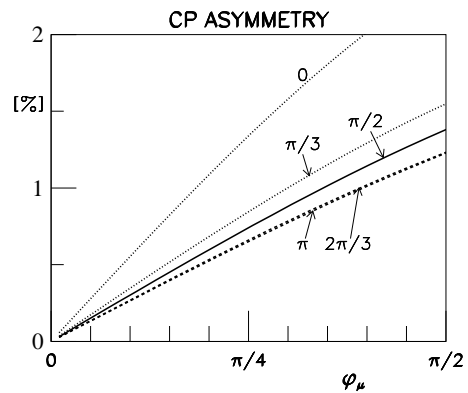}
\caption{
The ratio
(\ref{eq:obvservable-loop/tree}) for various 
values of the polar angle
$\theta$. The other parameters are fixed as
$\tan\beta = 5$, 
$\sqrt{s} = 800\text{ GeV}$.
\label{fig:asymm-800-cost-5} }
\end{center}
\end{figure}

The asymmetry demonstrates a smooth behaviour with respect to
the polar angle (see Fig.~\ref{fig:asymm-800-cost-5}).

\begin{figure}[htb]
\begin{center}
\includegraphics[width=7cm]{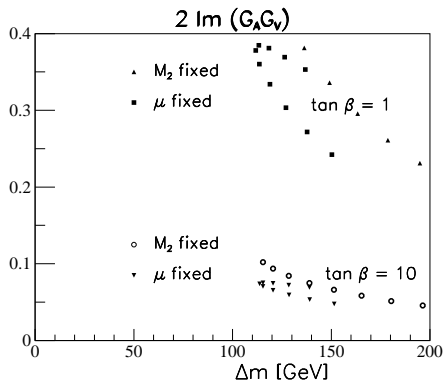}
\caption{
The couplings of
Eq.~(\ref{Eq:couplings}) vs.\ $\Delta m$, for two values of $\tan\beta$.
Some points are obtained with fixed $|\mu|=300~\text{GeV}$ and varying $M_2$,
others are obtained with fixed $M_2=200~\text{GeV}$ and varying $|\mu|$.
In all cases $\phi_\mu=\pi/2$.
\label{fig:couplings_vs_delta_m} }
\end{center}
\end{figure}

Since the effect somehow is due to the fact that the two chargino
mass states are different, one might think that it would vanish 
in the limit of equal chargino masses. This is not the case.
First of all, because of the finite $W$ mass, there is a minimum
splitting among the two chargino masses. The splitting would only 
vanish in the limit of $\mu\,M_2$ being real and negative, in which 
case there is no CP violation.
Secondly, these coupling constants do not correlate
very well with the mass difference, $\Delta m=m_{\chi_2}-m_{\chi_1}$.
This is illustrated in Fig.~\ref{fig:couplings_vs_delta_m},
where we show the quantity (\ref{Eq:couplings}) vs. $\Delta m$,
for the cases of fixed $M_2$ and fixed $|\mu|$, scanning over the other,
and two values of $\tan\beta$.

\section{Concluding remarks}
\label{sec:conclusion}

Since triangle diagrams have not been calculated, the results given in 
Figs.~\ref{fig:asymm_600}--\ref{fig:asymm-800-cost-5} are not to be 
seen as quantitative results, they are of a purely illustrative 
character. However, since the kinematic structure of the triangle 
diagrams is different from that of the box diagrams, when included,
these can not cancel the contributions of the box diagrams. Thus, we
conclude that the CP-violating asymmetry in the unpolarized cross
section is non-zero. Furthermore, we believe the effect, which depends
on the phases of both $\mu$ and $M_1$, to be of the
order of a percent.

To obtain the complete one-loop result for the observable
(\ref{eq:obvservable-definition}), one will also need to compute the
(triangle) loop corrections for each of the tree-level vertices
appearing in diagrams
Fig.~\ref{fig:tree} {\bf b}, {\bf c} and
{\bf d} (the Higgs coupling is negligible) assuming, of course, that
higgsino and U(1) gaugino mass parameters are complex.
Such a calculation would involve 40--60 diagrams (depending
on how one counts \cite{Diaz:2002rr}), and the
heavy-sneutrino limit does not lead to any obvious simplification.%
\footnote{We have recently been informed that such a calculation is in
progress by J. Kalinovski and K. Rolbiecki
\cite{Rolbiecki:2007se,Osland:2007ih}.}  In contrast to the box diagrams,
individual triangle diagrams are divergent. However, as was argued above,
since there is no contribution to the asymmetry
(\ref{eq:obvservable-definition}) at the tree level, they have to combine to a
finite quantity. What is interesting to note is that this calculation may
require the one-loop $\gamma\chi\chi$ vertex, absent at tree level. Indeed,
the $U(1)$ gauge invariance together with renormalizability protects the
photon from coupling with two fermions of different mass (see, e.g.,
\cite{'tHooft:1971fh}). However, the gauge invariance {\em alone} cannot
guarantee it: for example, the (non-renormalizable) vertex
\begin{equation}
\left( \bar\psi_{\chi_1}\sigma_{\mu \nu}\psi_{\chi_2} +
\bar\psi_{\chi_2}\sigma_{\mu \nu}\psi_{\chi_1} \right) F^{\mu\nu},
\end{equation}
where 
$\sigma_{\mu\nu} =
\frac{i}{4}\left[\gamma_{\mu},\gamma_{\nu}\right]$
and $F^{\mu\nu}$ is the field-strength tensor, provides such a
coupling, being explicitly gauge invariant.%
\footnote{
Couplings of this kind have been discussed in the context of neutrino
propagation, see, e.g.,
\cite{Okun:1986na}.
}
Hence, the (triangle) loop corrections to the
$\gamma\chi_i\chi_j$ vertex can in principle give a (UV-finite)
contribution to
$d\sigma_0$ and, possibly, to 
$d\sigma_0^{\rm odd}$. This has to be checked. Unfortunately, the
authors who recently reported complete one-loop calculations with real
couplings (CP-even case) do not comment on this
\cite{Blank:2000uc,Diaz:2002rr,Fritzsche:2004nf,Oller:2005xg}.

If the heavier chargino is considerably heavier than the lighter one,
it might be easier to observe CP-violation in the production of
equal-mass charginos, using polarized beams 
\cite{Moortgat-Pick:2005cw}, however the one-loop corrections to
polarized amplitudes require a separate study.

\begin{acknowledgments}
It is a pleasure to thank K.~Rolbiecki for pointing out an error
in the first version of this paper.
We are also grateful to A.~Bartl for communicating the results of
\cite{Bartl:2004xy} prior to publication and noticing an error in
our tree level results for polarized cross-section. 
Finally, we wish to
thank C.~Jarlskog and V.~Vereshagin for important communications and
discussions.
\end{acknowledgments}




\begin{thebibliography}{99}

\bibitem{MSSM}
  H.~P.~Nilles,
  Phys.\ Rept.\  {\bf 110}, 1 (1984);
  H.~E.~Haber and G.~L.~Kane,
  Phys.\ Rept.\  {\bf 117}, 75 (1985);

H.~Baer and X.~Tata, {\em Weak Scale Supersymmetry} (Cambridge 
University Press, Cambridge, England, 2006); see also Vol.~3 of
\cite{Weinberg_book}.
\bibitem{DreesBook}
  M.~Drees, R.~M.~Godbole, P.~Roy, {\em Theory and Phenomenology of
  Sparticles} (World Scientific, Singapore, 2004). 

\bibitem{Ellis:1982tk}
  J.~R.~Ellis, S.~Ferrara and D.~V.~Nanopoulos,
  Phys.\ Lett.\ B {\bf 114}, 231 (1982).
  W.~Buchmuller and D.~Wyler,
  Phys.\ Lett.\ B {\bf 121}, 321 (1983);
  J.~Polchinski and M.~B.~Wise,
  Phys.\ Lett.\ B {\bf 125}, 393 (1983);
  F.~del Aguila, M.~B.~Gavela, J.~A.~Grifols and A.~Mendez,
  Phys.\ Lett.\ B {\bf 126}, 71 (1983)
  [Erratum-ibid.\ B {\bf 129}, 473 (1983)].

\bibitem{Kizukuri:1992nj}
  Y.~Kizukuri and N.~Oshimo,
  Phys.\ Rev.\ D {\bf 46}, 3025 (1992).

\bibitem{MSSM_phases}
  T.~Ibrahim and P.~Nath,
  Phys.\ Rev.\ D {\bf 57}, 478 (1998)
  [Erratum-ibid.\ D {\bf 58}, 019901 (1998\ ERRAT, D60, 079903. 1999\ 
  ERRAT, D60, 119901. 1999)]
  [arXiv:hep-ph/9708456];
  M.~Brhlik, G.~J.~Good and G.~L.~Kane,
  Phys.\ Rev.\ D {\bf 59}, 115004 (1999)
  [arXiv:hep-ph/9810457];
  N.~Ghodbane, S.~Katsanevas, I.~Laktineh and J.~Rosiek,
  Nucl.\ Phys.\ B {\bf 647}, 190 (2002)
  [arXiv:hep-ph/0012031];
  A.~Bartl, W.~Majerotto, W.~Porod and D.~Wyler,
  Phys.\ Rev.\ D {\bf 68}, 053005 (2003)
  [arXiv:hep-ph/0306050];
  S.~Yaser Ayazi and Y.~Farzan,
  Phys.\ Rev.\  D {\bf 74}, 055008 (2006)
  [arXiv:hep-ph/0605272].

\bibitem{Accomando:1997wt}
  E.~Accomando {\it et al.}  [ECFA/DESY LC Physics Work\-ing Group],
  Phys.\ Rept.\  {\bf 299}, 1 (1998)
  [arXiv:hep-ph/9705442];
  J.~A.~Aguilar-Saavedra {\it et al.}  [ECFA/DESY LC Physics Working
  Group],
  arXiv:hep-ph/0106315;
  T.~Abe {\it et al.}  [American Linear Collider Working Group],
  in {\it Proc. of the APS/DPF/DPB Summer Study on the Future of
  Particle Physics (Snowmass 2001) } ed. N.~Graf, SLAC-R-570
  {\it Resource book for Snowmass 2001, 30 Jun - 21 Jul 2001,
  Snowmass, Colorado};
  G.~Weiglein {\it et al.}  [LHC/LC Study Group],
  Phys.\ Rept.\  {\bf 426}, 47 (2006)
  [arXiv:hep-ph/0410364].
  
\bibitem{Kizukuri:1993vh}
  Y.~Kizukuri and N.~Oshimo,
  Proc.\ Workshop on $e^+e^-$ Collisions at 500~GeV:
  The Physics Potential, Munich--Annecy--Hamburg 1993, DESY 93-123C,
  P. Zerwas (ed.);
  arXiv:hep-ph/9310224.

\bibitem{Choi:1998ei}
  S.~Y.~Choi, A.~Djouadi, H.~S.~Song and P.~M.~Zerwas,
  Eur.\ Phys.\ J.\  C {\bf 8}, 669 (1999)
  [arXiv:hep-ph/9812236];
  S.~Y.~Choi, M.~Guchait, J.~Kalinowski and P.~M.~Zerwas,  
  Phys.\ Lett.\  B {\bf 479}, 235 (2000)
  [arXiv:hep-ph/0001175];
  S.~Y.~Choi, A.~Djouadi, M.~Guchait, J.~Kalinowski, H.~S.~Song and P.~M.~Zerwas,
  Eur.\ Phys.\ J.\  C {\bf 14}, 535 (2000)
  [arXiv:hep-ph/0002033].

\bibitem{Bartl:2004vi}
  A.~Bartl, H.~Fraas, O.~Kittel and W.~Majerotto,
  Phys.\ Lett.\ B {\bf 598}, 76 (2004)
  [arXiv:hep-ph/0406309].

\bibitem{Yang:2002am}
  W.~M.~Yang and D.~S.~Du,
  Phys.\ Rev.\ D {\bf 67}, 055004 (2003)
  [arXiv:hep-ph/0211453].

\bibitem{Bartl:2004xy}
  A.~Bartl, K.~Hohenwarter-Sodek, T.~Kernreiter and H.~Rud,
  Eur.\ Phys.\ J.\ C {\bf 36}, 515 (2004)
  [arXiv:hep-ph/0403265].

\bibitem{Dass:1975mj}
  G.~V.~Dass and G.~G.~Ross,
  Phys.\ Lett.\ B {\bf 57}, 173 (1975);
  Nucl.\ Phys.\ B {\bf 118}, 284 (1977).

\bibitem{Ananthanarayan:2006yj}
  B.~Ananthanarayan and S.~D.~Rindani,
  Eur.\ Phys.\ J.\  C {\bf 46}, 705 (2006)
  [arXiv:hep-ph/0601199].

\bibitem{Hollik:2002mv}
  W.~Hollik, E.~Kraus, M.~Roth, C.~Rupp, K.~Sibold and D.~Stockinger,
  Nucl.\ Phys.\ B {\bf 639}, 3 (2002)
  [arXiv:hep-ph/0204350].

\bibitem{Ananthanarayan:2003wi}
  B.~Ananthanarayan and S.~D.~Rindani,
  Phys.\ Rev.\ D {\bf 70}, 036005 (2004)
  [arXiv:hep-ph/0309260].

\bibitem{Branco}
  G.~C.~Branco, L.~Lavoura, J.~P.~Silva, {\em CP Violation} (Oxford
  University Press, Oxford, England, 1999). 

\bibitem{Heinemeyer:2004gx}
  S.~Heinemeyer, W.~Hollik and G.~Weiglein,
  Phys.\ Rept.\  {\bf 425}, 265 (2006)
  [arXiv:hep-ph/0412214].

\bibitem{Allanach:2002nj}
 B.~C.~Allanach {\it et al.},
  in {\it Proc. of the APS/DPF/DPB Summer Study on the Future of
  Particle Physics 
  (Snowmass 2001) } ed. N.~Graf,
  {\it In the Proceedings of APS / DPF / DPB Summer Study on the
  Future of Particle Physics (Snowmass 2001), Snowmass, Colorado, 30
  June - 21 July 2001, pp P125}
  [arXiv:hep-ph/0202233].

\bibitem{Bjorken&Drell}
  J.~Bjorken and S.~Drell,
  {\em Quantum Quantum Mechanics} (McGraw-Hill, 1964)

\bibitem{Haber:1994pe}
  H.~E.~Haber,
  in Proceedings of the XXI SLAC Summer Institute on
  Particle Physics ``Spin Structure in High Energy Processes'',
  edited by L.~DePorcel and Ch.~Dunwoodie, Springfield, 1994;
  arXiv:hep-ph/9405376.

\bibitem{Weinberg_book}
  S.~Weinberg, {\em The Quantum Theory of Fields} (Cambridge
  University Press, Cambridge, England, 2000), Vols. 1--3.

\bibitem{FeinWein}
  G.~Feinberg and S.~Weinberg, Nuovo Cimento {\bf 14}, 571 (1959).

\bibitem{Blank:2000uc}
  T.~Blank and W.~Hollik,
  in {\em 2nd ECFA/DESY Study 1998--2001};
  arXiv:hep-ph/0011092.

\bibitem{Diaz:2002rr}
  M.~A.~Diaz and D.~A.~Ross,
  JHEP {\bf 0106}, 001 (2001)
  [arXiv:hep-ph/0103309];
  arXiv:hep-ph/0205257.

\bibitem{Fritzsche:2004nf}
  T.~Fritzsche and W.~Hollik,
  Nucl.\ Phys.\ Proc.\ Suppl.\  {\bf 135}, 102 (2004)
  [arXiv:hep-ph/0407095].

\bibitem{Oller:2005xg}
  W.~Oller, H.~Eberl and W.~Majerotto,
  Phys.\ Rev.\  D {\bf 71}, 115002 (2005)
  [arXiv:hep-ph/0504109].

\bibitem{Kileng}
  B.~Kileng, dr.\ scient.\ thesis, University of Bergen, 1994;
  some sign inconsistencies are corrected.

\bibitem{Rosiek:1989rs}
  J.~Rosiek,
  Phys.\ Rev.\ D {\bf 41}, 3464 (1990);
  the version with the most recent corrections is available at:
  {\tt http:// www.fuw.edu.pl/$\sim$rosiek/physics/prd41.html}

\bibitem{'tHooft:1971fh}
  G.~'t Hooft,
  Nucl.\ Phys.\ B {\bf 33}, 173 (1971).

\bibitem{Fujikawa:1972fe}
  K.~Fujikawa, B.~W.~Lee and A.~I.~Sanda,
  Phys.\ Rev.\ D {\bf 6}, 2923 (1972).

\bibitem{Non-linearGauge}
  Non-linear gauges are discussed, e.g., in
  K.~Fujikawa,
  Phys.\ Rev.\ D {\bf 7}, 393 (1973);
  G.~Keller and D.~Wyler,
  Nucl.\ Phys.\ B {\bf 274}, 410 (1986);
  M.~B.~Gavela, G.~Girardi, C.~Malleville and P.~Sorba,
  Nucl.\ Phys.\ B {\bf 193}, 257 (1981).

\bibitem{Allanach:2003eb}
  B.~C.~Allanach, A.~Dedes and H.~K.~Dreiner,
  Phys.\ Rev.\ D {\bf 69}, 115002 (2004)
  [Erratum-ibid.\ D {\bf 72}, 079902 (2005)]
  [arXiv:hep-ph/0309196], and references therein.

\bibitem{Denner:1992vz}
  A.~Denner, H.~Eck, O.~Hahn and J.~Kublbeck,
  Nucl.\ Phys.\ B {\bf 387}, 467 (1992);
  Phys.\ Lett.\ B {\bf 291}, 278 (1992).

\bibitem{Vermaseren:1994je}
  J.~A.~M.~Vermaseren,
  Comput.\ Phys.\ Commun.\  {\bf 83}, 45 (1994);
  D.~Binosi and L.~Theussl,
  Comput.\ Phys.\ Commun.\  {\bf 161}, 76 (2004)
  [arXiv:hep-ph/0309015].

\bibitem{Passarino:1978jh}
  G.~Passarino and M.~J.~G.~Veltman,
  Nucl.\ Phys.\ B {\bf 160}, 151 (1979);
  G.~'t Hooft and M.~J.~G.~Veltman,
  Nucl.\ Phys.\ B {\bf 153}, 365 (1979).

\bibitem{PhD}
  A.~Vereshagin, PhD thesis (University of Bergen, Norway, in
  preparation).

\bibitem{Hahn:1998yk}
  T.~Hahn and M.~Perez-Victoria,
  Comput.\ Phys.\ Commun.\  {\bf 118}, 153 (1999)
  [arXiv:hep-ph/9807565].\\
  See also http://www.feynarts.de/looptools/

\bibitem{vanOldenborgh:1989wn}
  G.~J.~van Oldenborgh and J.~A.~Vermaseren,
  Z.\ Phys.\ C {\bf 46}, 425 (1990).

\bibitem{Moortgat-Pick:2005cw}
  G.~A.~Moortgat-Pick {\it et al.},
  arXiv:hep-ph/0507011.

\bibitem{Rolbiecki:2007se}
  K.~Rolbiecki and J.~Kalinowski,
  arXiv:0709.2994 [hep-ph].

\bibitem{Osland:2007ih}
  P.~Osland, J.~Kalinowski, K.~Rolbiecki and A.~Vereshagin,
  arXiv:0709.3358 [hep-ph].

\bibitem{Okun:1986na}
  L.~B.~Okun, M.~B.~Voloshin and M.~I.~Vysotsky,
  Sov.\ Phys.\ JETP {\bf 64}, 446 (1986)
  [Zh.\ Eksp.\ Teor.\ Fiz.\  {\bf 91}, 754 (1986)];
  Sov.\ J.\ Nucl.\ Phys.\  {\bf 44}, 440 (1986)
  [Yad.\ Fiz.\  {\bf 44}, 677 (1986)].

\end{thebibliography}
\end{document}